\documentclass[12pt]{article}
\newcommand{\ba}{\begin{array}}
\newcommand{\ea}{\end{array}}
\newcommand{\no}{\nonumber}
\newcommand{\nn}{\nonumber\\}

\newcommand{\R}{\mathbb{R}}
\newcommand{\C}{\mathbb{C}}

\newcommand{\cD}{{\cal D}}

\newcommand{\cH}{{\cal H}}

\newcommand{\cN}{{\cal N}}

\newcommand{\cP}{{\cal P}}

\newcommand{\x}{\xi}

\newcommand{\om}{\omega}

\newcommand{\del}{\partial}
\newcommand{\vvert}{\Big{\vert}}
\newcommand{\bra}{\langle}
\newcommand{\ket}{\rangle}

\newcommand{\Bra}{\Big{\langle}}
\newcommand{\Ket}{\Big{\rangle}}

\newcommand{\Ker}{{\rm Ker}\,}

\newcommand{\pr}{\prime}

\newcommand{\rar}{\rightarrow}

\newcommand{\dar}{\downarrow}

\newcommand{\longr}{\longrightarrow}
\newcommand{\longl}{\longleftarrow}
\newcommand{\longlr}{\longleftrightarrow}

\newcommand{\map}{\mapsto}

\newcommand{\ti}{\times}

\newcommand{\ot}{\otimes}

\newcommand{\st}{\stackrel}
\newcommand{\unb}{\underbrace}

\newcommand{\lab}{\label}
\newcommand{\fr}{\frac}
\newcommand{\half}{\frac{1}{2}}

\newcommand{\dis}{\displaystyle}
\newcommand{\mn}{{\mu\nu}}
\newcommand{\eb}{\bar{e}}

\newcommand{\zb}{\bar{z}}
\newcommand{\cDb}{\bar{\cD}}

\newcommand{\ah}{\hat{a}}
\newcommand{\fh}{\hat{f}}
\newcommand{\gh}{\hat{g}}

\newcommand{\nh}{{\hat{\nu}}}

\newcommand{\xh}{\hat{x}}
\newcommand{\zh}{\hat{z}}

\newcommand{\Dh}{\hat{D}}
\newcommand{\Fh}{\hat{F}}
\newcommand{\Ph}{\hat{P}}

\newcommand{\Uh}{\hat{U}}
\newcommand{\Vh}{\hat{V}}

\newcommand{\zbh}{\hat{\bar{z}}}

\newcommand{\delh}{\hat{\partial}}

\usepackage{graphicx}
\usepackage{amssymb}
\usepackage{amsmath}
\usepackage{color}

\makeatletter
\@addtoreset{equation}{section}

\makeatother

\begin{document}

\begin{titlepage}
\null
\begin{flushright}
%-/-
%\\
arXiv:yymm.nnnn
\\
November, 2013
\end{flushright}

\vskip 2cm
\begin{center}

{\Large \bf ADHM Construction of Noncommutative Instantons}

\vskip 2cm
\normalsize

{\large Masashi Hamanaka\footnote{E-mail: hamanaka@math.nagoya-u.ac.jp}
 and
 Toshio Nakatsu\footnote{E-mail: nakatsu@mpg.setsunan.ac.jp}}

\vskip 1.5cm

        $~^1${\it Graduate School of Mathematics, Nagoya University,\\
                     Chikusa-ku, Nagoya, 464-8602, JAPAN}
\vskip 0.5cm

        $~^2${\it Institute for Fundamental Sciences, Setsunan University,\\
             17-8 Ikeda Nakamachi, Neyagawa, Osaka, 572-8508, JAPAN}

\vskip 1.5cm

{\bf \large Abstract}

\end{center}

We discuss the Atiyah-Drinfeld-Hitchin-Manin (ADHM) construction of $U(N)$ instantons 
in noncommutative (NC) space and prove the one-to-one correspondence between moduli spaces 
of the noncommutative instantons and the ADHM data,  
together with an origin of the instanton number for $U(1)$. 
We also give a derivation of the ADHM construction from the viewpoint of 
the Nahm transformation of instantons on four-tori. 
This article is a composite version of 
\cite{HaNa_JPCS} and \cite{HaNa_FMC}.
\end{titlepage}

\clearpage

\baselineskip 6mm

\section{Introduction}

Anti-self-dual (ASD) Yang-Mills equation 
and the solutions have been studied from the several viewpoints of mathematical physics, 
particularly, integrable systems, geometry and field theories.  
The integrable structure is well understood by using the idea of the
twistor theory founded by R. Penrose (see, e.g., \cite{Penrose}). 
In particular, it was applied to construct the exact solutions, 
where the self-duality of gauge fields on $S^4$ is converted to 
the holomorphy of complex vector bundles on $\mathbb{C} P_3$. 
It also reveals a relationship to the Riemann-Hilbert problem and 
has made much progress with integrable systems (see e.g. \cite{WaWe}). 
It was first observed by R. Ward, 
now called the {Ward conjecture}, 
that lower-dimensional soliton equations such as the KdV equation 
are obtained from the anti-self-dual Yang-Mills equation by dimensional reduction.  
The twistor theoretical approach matches with the reductions and leads a geometrical understanding 
of the integrability (see e.g. \cite{MaWo}).

Instantons are finite-action solutions of the anti-self-dual Yang-Mills equation.  
Hence they are exact solutions of classical Yang-Mills theories. 
Instantons can reveal non-perturbative aspects of the quantum theories, 
since the path-integrations, formulating the quantum theories, 
reduce to finite-dimensional integrations over the instanton moduli spaces. 
To obtain the exact solutions, 
the Atiyah-Drinfeld-Hitchin-Manin (ADHM) construction \cite{ADHM} is a powerful method. 
In particular, the moduli space is mapped to the set of quadruple matrices via the construction.  
The aforementioned integration is converted to an integration over the matrices \cite{DHKM}. 

Together with the use of noncommutative instantons \cite{NeSc}, 
it can be achieved \cite{Nekrasov4, NaYo} by applying a localization formula. 
In the procedures, various formulas and relations of the ADHM construction are required. 
Hence it is worthwhile to elucidate the noncommutative ADHM construction 
together with the group actions and to present all the ingredients in the construction explicitly. 

In our paper \cite{HaNa}, 
we discuss the ADHM construction of $U(N)$ instantons in noncommutative spaces 
and prove one-to-one correspondence between moduli spaces of the noncommutative instantons and the ADHM data. 
We also argue group actions (especially torus actions) on noncommutative instantons. 
The present paper is a brief report of it. % (See also \cite{HaNa2}.)

\section{Noncommutative Spaces and Field Theories}

Noncommutative (NC) space is a space which coordinate ring is noncommutative. 
Let $x^{\mu}$ be the spacial coordinates. 
The noncommutativity is expressed by the following commutation relations. 
\begin{eqnarray}
\label{nc_coord} 
[ x^\mu, x^\nu ] = i \theta^{\mu \nu},
\end{eqnarray}
where $\theta^{\mu\nu}$  is a real antisymmetric tensor 
and called {\it noncommutative parameters}. 
When $\theta^{\mu \nu}$ vanishes identically, 
the coordinate ring is commutative 
and the underlying space reduces to a commutative one.  
The commutation relations (\ref{nc_coord}), 
like the canonical commutation relations in quantum mechanics, 
lead to ``space-space uncertainty relation.'' 
Singularities in commutative space could resolve in noncommutative space thereby. 
This is one of the prominent features of field theories on noncommutative space,  
NC field theories, for short, 
and yields various new physical objects such as $U(1)$ instantons. 
We will examine NC field theories on noncommutative Euclidean spaces.
In four-dimensional Euclidean space, 
the antisymmetric tensor takes the canonical form as 
\begin{eqnarray}
\label{can_nc_coord}
\theta^{\mn} 
= \left( \ba{cc|cc}
           0 & 
-\theta_1 &         0 &         0 \\
\theta_1 &        0 &         0 &         0 \\\hline
           0 &        0 &         0 &  -\theta_2 \\
           0 &        0 & \theta_2 &         0 \\
  \ea \right), 
\end{eqnarray}
where $\theta_1$ and $\theta_2$ are real numbers. 
The condition that $\theta^{\mn}$ is self-dual (anti-self-dual) 
corresponds to $\theta_1-\theta_2=0$ ($\theta_1+\theta_2=0$) in the canonical form.  
In this section, we describe noncommutative gauge theories in two different manners. 
One is the star-product formalism and the other is the operator formalism. 
These descriptions turn to be equivalent by the Weyl transformation.

\subsection{Star-product formalism}

The Moyal star-product $f \star g$, 
where $f = f(x)$ and $g = g(x)$ are functions over $\mathbb{R}^4$, is 
a noncommutative associative product which is given by 
\begin{eqnarray}
\label{Moyal_product}
f \star g ( x ) 
=
\left. \exp{ \left( \frac{i}{2} \theta^{\mu \nu} 
                       \partial^{ ( x^\prime ) }_\mu 
                       \partial^{ ( x^{ \prime \prime } ) }_\nu \right) }
       f( x^\prime ) g( x^{ \prime \prime } ) 
\right|_{ x^{\prime} = x^{ \prime \prime } = x }. 
\end{eqnarray} 
The right-hand side can be expanded in powers of $\theta$ as
\begin{eqnarray}
\label{expansion_Moyal_product}
f \star g ( x ) 
= f(x) g(x) + \frac{i}{2} \theta^{\mu \nu} \partial_\mu f( x ) \partial_\nu g(x)
+ O  ( \theta^2 ).
\end{eqnarray}
Endowed with the star-product,
ring of functions over $\mathbb{R}^4$ becomes a noncommutative ring. 
In particular,  
the spatial coordinates satisfy 
$[ x^\mu, x^\nu ]_\star = i \theta^{ \mu \nu }$, 
where $[ f, \,g ]_\star \equiv f \star g - g \star f$. 
In the limit $\theta^{\mu\nu}\rar 0$, 
the product $f \star g$ becomes the standard product $f \cdot g$, 
as seen from (\ref{expansion_Moyal_product}). 
This means that NC field theories are a deformation from commutative field theories. 
We may call them {\it NC-deformed theories}.

In NC gauge theories, gauge transformation of the gauge potential $A$, 
often called the star gauge transformation \cite{Szabo}, is of the form 
\begin{eqnarray}
\label{star_gauge_transform}
A_\mu  
\,\,\mapsto\,\,   
g^{-1} \star A_\mu \star g + g^{-1} \star \del_\mu g,
\end{eqnarray}
where $g = g (x)$ takes values in the gauge group $G$. 
The field strength of $A$ is 
\begin{eqnarray}
\label{star_F} 
F^\star_{ \mu \nu } 
= \partial_\mu A_\nu - \partial_\nu A_\mu + [ \,A_\mu, \,A_\nu ]_\star .   
\end{eqnarray}
The field strength (\ref{star_F}) transforms covariantly 
under the gauge transformation (\ref{star_gauge_transform}) 
as $F^{\star} \mapsto g^{-1} \star F^\star \star g$. 
For the covariant gauge transform, 
the term $[ A_\mu, \,A_\nu ]_\star$ in (\ref{star_F}) 
is needed even in the case of $G = U(1)$.

NC-deformed version of anti-self-dual Yang-Mills equation can be presented in the Lax formulation.
Let $G = U(N)$. 
$D = d + A$ is the covariant derivative in the Yang-Mills theory. 
We introduce two linear operators $L$ and $M$ as 
\begin{eqnarray}
L = D_{z_2} - \lambda D_{\bar{z}_1}, 
&&
M = D_{z_1} + \lambda D_{\bar{z}_2},
\end{eqnarray} 
where $z_1 = x^2 + ix^1$ and $z_2 = x^4 + ix^3$ are complex coordinates, 
and $\lambda$ is an auxiliary parameter (the spectral parameter). 
Consider the following linear equations.  
\begin{eqnarray}
L \star \Psi( x ; \lambda ) = 0, 
&&
M \star \Psi( x ; \lambda ) = 0,
\label{lin_asdym}
\end{eqnarray}
The compatibility condition of (\ref{lin_asdym}) is $[ L, \,M ]_\star = 0$. 
Since $[ L, \,M ]_\star$ is a quadratic polynomial of $\lambda$,  
the compatibility condition is nothing but 
vanishing each coefficients of the polynomial and becomes 
\begin{eqnarray}
F^\star_{ z_1 \bar{z}_1 } + F^\star_{ z_2 \bar{z}_2 } = 0, 
&& 
F^\star_{ z_1 z_2 } = 0.
\label{asdym}
\end{eqnarray}
This is actually the anti-self-dual equation 
$F^\star_{\mn}=-(*F^\star)_{\mn}$, 
where the Hodge $*$-operator is given by 
$(*F^\star)_{\mn} = (1/2) \epsilon_{\mn \rho \sigma}F^{\star \rho \sigma}$.

The Lax representation of the anti-self-dual Yang-Mills equation 
with the linear operators in specific forms, gives lower-dimensional integrable equations. 
Several examples of such reductions for NC-deformed anti-self-dual Yang-Mills equation 
can be found in \cite{Hamanaka2}.
The Riemann-Hilbert problem associated to the linear system
\eqref{lin_asdym} is discussed in \cite{GHN, Takasaki}.

\subsection{Operator formalism}

Commutation relation (\ref{nc_coord}) hints that 
the coordinates $x^{\mu}$ are hermitian operators acting on a Hilbert space. 
To distinguish the operators from the previous ones, 
instead of $f$, the symbol $\hat{f}$ is used to represent the operator.

We outline the formalism in the case of a noncommutative two-plane 
whose coordinates $\xh^1$ and $\xh^2$ satisfy 
$[ \xh^1, \xh^2] = i \theta$ ($\theta>0$). 
Let $\zh = \xh^1 + i \xh^2$ and $\zbh = \xh^1 - i \xh^2$.  
These operators are normalized to obtain 
\begin{eqnarray}
\ah = \fr{1}{ \sqrt{2 \theta} } \zh, 
&& 
\ah^\dagger = \fr{1}{ \sqrt{2 \theta} }\zbh.
\end{eqnarray} 
The commutation relation becomes Heisenberg's commutation relation. 
\begin{eqnarray}
\label{heisenberg}
[ \ah, \,\ah^\dagger ] = 1.
\end{eqnarray}
The Fock vacuum $\vert 0 \ket$ satisfies the condition $\ah\vert 0\ket=0$.
Let $\cH$ be the Fock space built on $\vert 0 \ket$. We have   
\begin{eqnarray}
\label{fock}
\cH = \bigoplus_{n=0}^{+\infty} \,\C \vert n \ket, 
\end{eqnarray}
where 
$\vert n \ket = \left\{ ( \ah^\dagger )^n / \sqrt{n!} \right\} \vert 0 \ket$ ($n=0, 1, 2 ,\dots$). 
$\hat{f} = \fh( \xh^1, \xh^2 )$ is an operator on $\cH$. 
In the occupation number basis (\ref{fock}), 
it can be written as a semi-infinite matrix of the form
\begin{eqnarray}
\fh = \sum_{m,n=0}^{ + \infty } f_{mn} \vert m \ket \bra n \vert.
\end{eqnarray}
If $\hat{f}$ possesses the rotational symmetry, 
it commutes with the number operator 
$\nh = \ah^\dagger \ah \sim (\xh^1)^2 + (\xh^2)^2$ 
and therefore is a diagonal matrix of the form 
\begin{eqnarray}
\fh = \sum_{n=0}^{+ \infty } f_{n} \vert n \ket \bra n \vert.
\end{eqnarray}

Let $\hat{P}_k$ be a projection of $\cH$ onto a $k$-dimensional subspace.  
With such a projection operator we associate the {\it shift operator}.
It is an operator $\hat{U}_k$ which satisfies 
\begin{eqnarray}
\label{partial_iso}
\Uh_k \Uh_k^\dagger = 1, ~~~\Uh_k^\dagger \Uh_k = 1 - \Ph_k.
\end{eqnarray} 
Shift operators play a central role in the construction of noncommutative solitons. 
In two-dimensions, for instance, 
$\hat{P}_k = \sum_{p=0}^{k-1} \vert p \ket \bra p \vert$ 
is a projection operator of rank $k$ and the shift operator takes the following form.  
\begin{eqnarray}
\Uh_k = \sum_{n=0}^{+ \infty } \vert n \ket \bra n+k \vert. 
\end{eqnarray}

In four-dimensions, 
using the complex coordinates 
$\hat{z}_1 = \hat{x}^2 + i \hat{x}^1$ and $\hat{z}_2 = \hat{x}^4 + i \hat{x}^3$
together with the canonical form \eqref{can_nc_coord} of $\theta^{\mn}$, 
the noncommutativity can be expressed as   
\begin{eqnarray}
\label{nc_cpx_coord} 
[ \hat{z}_1, \hat{\zb}_1 ] = 2 \theta_1, ~~~[ \hat{z}_2, \hat{\zb}_2 ] = 2 \theta_2. 
\end{eqnarray} 
Each pairs of the complex coordinate and its complex conjugation in (\ref{nc_cpx_coord}) 
provides annihilation and creation operators of a harmonic oscillator.
For instance, when both $\theta_{j = 1, 2}$ are negative numbers, 
$( 1/{\sqrt{ -2 \theta_j } }) \hat{\bar{z}}_j$ are the annihilation operators $a_j$, 
while $( 1/{ \sqrt{ -2 \theta_j } }){\hat{z}}_j$ are the creation operators $a_j^{\dagger}$. 
Let $\cH_j$ be the Fock spaces of each harmonic oscillators.  
The Hilbert space in four-dimensions is the tensor product $\cH = \cH_1 \otimes \cH_2$. 
Similarly to (\ref{fock}), we have 
\begin{eqnarray}
\label{cH_4D}
\cH = \bigoplus_{n_1, n_2 = 0}^{+\infty} \,\C \vert n_1, n_2 \ket, 
\end{eqnarray}
where $\vert n_1, n_2 \rangle = \vert n_1 \rangle \otimes \vert n_2 \rangle ~(n_1, n_2 = 0, 1, \dots)$ 
are the occupation number basis.

$\hat{f} = \hat{f}( \hat{x} )$ is an operator on $\cH$.  
In the basis (\ref{cH_4D}), 
it can be written as an infinite matrix of the form  
\begin{eqnarray}
\label{f_hat}
\fh( \xh ) &=& \hspace{-6mm} \sum_{m_1, m_2, n_1, n_2 = 0}^{+\infty} 
\hspace{-5mm} f_{m_1, m_2, n_1, n_2} \vert m_1, m_2 \ket \bra n_1, n_2 \vert.
\end{eqnarray}
Noncommutative anti-self-dual Yang-Mills equation takes the following form 
in the operator formalism.  
\begin{eqnarray}
\label{bps_instanton}
\Fh_{ z_1 \zb_1 } + \Fh_{ z_2 \zb_2 } = 0, ~~~\Fh_{ z_1 z_2 } = 0.
\end{eqnarray}

\subsection{Weyl Transformation}

The two descriptions of NC field theories are equivalent. 
The first description can be converted to the second by applying the Weyl transformation, 
and the inverse is achieved by the inverse Weyl transformation. 
Let us explain these transformations briefly in the case of the noncommutative two-plane. 
$f = f( x_1, x_2 )$ in the star-product description is converted to an operator 
$\hat{f} = \hat{f}( \hat{x}_1, \hat{x}_2 )$ by the following Weyl transformation. 
\begin{eqnarray}
\label{weyl1}
\fh ( \xh^1, \xh^2) 
= \fr{1}{( 2 \pi )^2 } \int dk_1 dk_2
      \,\,\widetilde{f}( k_1, k_2 ) \,e^{ -i( k_1 \xh^1 + k_2 \xh^2 )},
\end{eqnarray}
where
\begin{eqnarray}
\label{weyl2}
\widetilde{f} ( k_1, k_2 ) 
= \int dx^1 dx^2 \,\,f( x^1, x^2 ) \,e^{i( k_1 x^1 + k_2 x^2 )}.
\end{eqnarray} 
The above transform is a composition of the Fourier 
and the inverse Fourier transforms. 
Under the inverse transform, 
the commutative coordinates $x^1, x^2$ of the integral kernel 
are replaced with the noncommutative coordinates $\xh^1, \xh^2$ 
and thereby the Fourier transform $\widetilde{f}$ is mapped to the operator $\fh$. 
The inverse Weyl transformation is given directly by
\begin{eqnarray}
f(x^1,x^2) 
= \int dk_2~e^{- \frac{ ik_2 x^2}{\theta}} 
     \Bra x^1 + \fr{k_2}{2} \vvert  
            \fh( \xh^1, \xh^2 )
     \vvert x^1 - \fr{k_2}{2} \Ket.
\end{eqnarray}

The Weyl transformation, as well as the inverse transformation, preserves the products. 
\begin{eqnarray}
\label{algebra_hom_1}
\widehat{f \star g} = \fh \cdot \gh.
\end{eqnarray}
These transformations also commute with 
operations of differentiation and integration formulated in each descriptions.  
Thus, two equations (\ref{asdym}) and (\ref{bps_instanton}) are equivalent. 
They are representations of the noncommutative anti-self-dual Yang-Mills equation in each descriptions.  
The correspondence is listed in the table. 
For more details on noncommutative field theories, 
see e.g. \cite{Chu, DoNe, HaNa, Harvey2, KoSc, SeWi, Szabo}.

%%%%%%%%%%%%%%%%%%%%%%%%%
\begin{center}
%\caption{}
\begin{tabular}{|c|c|c|} \hline
Star-product formalism  & $\leftarrow$ Weyl transformation $\rightarrow$ & Operator formalism 
\\ \hline\hline 
     $f(x^1,x^2)$       &      Fields    &  $\dis \fh(\xh^1,\xh^2) = \sum_{m,n=0}^{+\infty}f_{mn}\vert m\ket\bra n\vert$
\\\hline
star product $f\star g$  & Product  & matrix multiplication  $\hat{f}\cdot \hat{g}$ 
\\
\hline
${[x^\mu,x^\nu]}_\star=i\theta^{\mu\nu}$ & Noncommutativity & $[\hat{x}^\mu,\hat{x}^\nu]=i\theta^{\mu\nu}$
\\\hline
$\del_{\mu} f$  & Differentiation  
& $\del_{\mu}\fh:= [\unb{-i(\theta^{-1})_{\mu\nu}\xh^{\nu}}_{=:~\delh_{\mu}},\fh]$
\\
\hline
$\dis\int dx^1dx^2~f(x^1,x^2)$ & Integration & $2\pi\theta{\mbox{Tr}}_{\cH}\fh(\xh^1,\xh^2)$
\\\hline
$\ba{c} \dis\sqrt{\fr{n!}{m!}}
\left(2r^2/\theta\right)^{\fr{m-n}{2}}e^{i(m-n)\varphi}\times\\
2(-1)^nL_n^{m-n}(2r^2/\theta)e^{-\fr{r^2}{\theta}}
\ea$
& $\ba{c} {\mbox{Matrix elements}}\\ (m\geq n)\ea$  & $\vert n\ket\bra m\vert$
\\
  &  &
\\[-3mm]
$\dis 2(-1)^nL_n(2r^2/\theta)e^{-\fr{r^2}{\theta}}$ & & $\vert n\ket\bra n\vert$
\\\hline
\end{tabular}
\end{center}
%%%%%%%%%%%%%%%%%%%
$( r, \,\varphi )$ is the polar coordinates.  
$\displaystyle L_n^\alpha (x)$ and $\displaystyle L_n (x) = L_n^0 (x)$ $( n = 0, 1, \dots )$ 
are the Laguerre polynomials.

\section{Nahm Transformation and Origin of the ADHM Duality}

The ADHM construction is a descendant of the twistor theory founded by R. Penrose. 
 Twistor theoretical idea was first applied by R. Ward to instantons. 
It converts the self-duality of gauge fields on $S^4$ 
to the holomorphy of complex vector bundles on $\mathbb{C} P_3$ (see e.g. \cite{WaWe}), 
where the construction reduces to an algebro-geometric one,  
and finally Atiyah, Drinfeld, Hitchin and Manin obtained 
an algebraic method to generate {\it all} instanton solutions on $S^4$ \cite{ADHM}.
(Instantons on $S^4$ are equivalent to those on $\mathbb{R}^4$ by the conformal invariance and Uhlenbeck's theorem.)
The construction is based on one-to-one correspondence between the moduli space of instantons and that of the ADHM data. 
The correspondence is thereafter called the {\it ADHM duality}.

 It was also applied to the case of the BPS monopoles by W. Nahm, 
which is called the {\it ADHMN construction} or the {\it Nahm construction}. 
Schenk \cite{Schenk} and Braam and van Baal \cite{BrvB}, 
extracting the Fourier transform, known as the Nahm transformation, from the construction 
and applying to instantons on four-torus, obtained the duality between instantons on four-torus and its dual torus, 
which is also known as the {\it Nahm duality}. 
Nahm transformations of explicit ASD gauge fields are performed in e.g. \cite{ACWTW, HaKa, Hori, vanBaal}. 
For surveys of the Nahm transformation, see e.g. \cite{BBR, DoKr, Jardim}.

The ADHM duality is also understood as a version of the Nahm duality.
The ADHM duality in this line was first outlined in \cite{CoGo3} and made rigorously in \cite{KrNa} 
with a generalization to ALE spaces.
(A beautiful treatment can be also found in \cite{DoKr}.)
Noncommutative version of the ADHM duality is discussed in \cite{HaNa, MaSa, Nekrasov3}.
%Nahm transformation is a duality transformation (one-to-one mapping) between
%the instanton moduli space on a four-torus $T^4$ with $G=U(N),~C_2=k$
%and that on the dual torus $\widetilde{T}^4$ with $\widetilde{G}=U(k),~C_2=N$.
%This situation is realized as D0-D4 brane systems where the D4-branes wrap on $T^4$.
%We can take T-duality transformation in the four directions where the D4-brane lie, which is just the Nahm transformation.
We start this section by giving a brief review on the Nahm transformation 
and then argue the ADHM/Nahm duality by taking certain limits.

\subsection{Poincar\'e Line Bundle}

%%In order to set up the stage, we introduce the Poincar\'e line bundle first.

The Poincar\'e line bundle plays a central role in the Nahm duality. 

The Poincar\'e line bundle $\cP$ is a $U(1)$ line bundle over the product space 
$T^4 \times \widetilde{T}^4$, where $T^4$ is a four-torus and $\widetilde{T}^4$ is the dual four-torus. 
The four torus is realized as $T^4 = \mathbb{R}^4 / \Lambda$, 
where $\Lambda$ is a maximal lattice in $\mathbb{R}^4$. 
The dual four-torus is $\widetilde{T}^4 = \mathbb{R}^{4 *} / 2 \pi \Lambda^*$, 
where $\mathbb{R}^{4 *}$ is the dual vector space and $\Lambda^*$ is the dual lattice.   
We use $x^\mu$ and $\x_\mu$ as the coordinates of $\R^4$ and $\R^{4*}$ 
with the dual pairing $\x \cdot x = \x_\mu x^\mu$. 
%%Let  $\Lambda$ be a rank-four lattice of $\R^4$.
%%Then a four-torus $T^4$ and the dual torus $\widetilde{T}^4$ are given as follows:
%%\begin{eqnarray}
%%T^4:=\R^4/\Lambda,~~~\widetilde{T}^4:=\R^{4*}/2\pi\Lambda^*,
%%\end{eqnarray}
%%where $\R^{4*}$ is the dual vector space of $\R^4$ and $\Lambda^*$ is the dual lattice of $\Lambda$: 
%%In this section, the dot ``$\cdot$ '' denotes 
%%the inner product of the elements of $\R^4$ and $\R^{4*}$.
%%\begin{eqnarray}
%%\Lambda^*:=\left\{\mu\in\R^{4*}~\vert~ \mu\cdot\lambda\in\Z,\all\lambda\in
%%\Lambda\right\}.
%%\end{eqnarray}
%%In this section, the dot ``$\cdot$ '' denotes the inner product of the elements of $\R^4$ and $\R^{4*}$.
%
%
%Roughly speaking, the torus and the dual torus have
%the opposite size to each other: $(\mbox{vol}\,T^4)
%\cdot(\mbox{vol}\,\widetilde{T}^4)=(2\pi)^4$.
%
%
%%The coordinates of $\R^4$ and $\R^{4*}$ are represented as $x^\mu$ and $\x_\mu$, respectively. 
The line bundle $\cP$ is equipped with a $U(1)$ potential of the form 
\begin{eqnarray}
\label{omega(x,xi)}
\omega (x, \x ) = i \x_\mu dx^\mu .
\end{eqnarray} 
Note that, taking another pair $( x', \x')$, where $x' \in x + \Lambda$ and $\xi' \in \xi + 2 \pi \Lambda^*$, 
amounts to $U(1)$ gauge transformation. 
Actually, it is apparent to see that, 
for $x' = x + \lambda$ $( \lambda \in \Lambda)$ and $\x' = \x + 2\pi \mu$ $( \mu \in \Lambda^* )$, 
two gauge potentials $\omega( x', \x' )$ and $\omega( x, \xi)$ are gauge equivalent.   
\begin{eqnarray}
\lab{omega_gauge}
\omega( x + \lambda, \x + 2\pi \mu ) = \omega( x, \x ) + g^{-1}(x, \x) dg(x, \x), 
\end{eqnarray}
where $g(x, \x) = e^{2\pi i\mu \cdot x}$ is single-valued on $T^4 \times \widetilde{T}^4$.  

The field strength $\Omega = d \omega$ becomes constant.
\begin{eqnarray}
\Omega( x, \x ) = i d\x_\mu \wedge dx^\mu.
\end{eqnarray}

The following form of the gauge potential is also used conveniently.  
\begin{eqnarray}
\label{omega'(x,xi)}
\omega^\pr (x,\x) = -i x^\mu d\x_\mu.
\end{eqnarray}
The above form can be obtained from (\ref{omega(x,xi)}) 
by the gauge transformation $\exp(- i\x\cdot x)$ on $\R^4\ti\R^{4*}$. 
\begin{eqnarray}
\label{omega_omega'}
\omega^\pr (x,\x) = \om(x,\x) +e^{i\x \cdot x}de^{-i\x \cdot x}. 
\end{eqnarray}

Let $\pi$, $\tilde{\pi}$ be natural projections from $T^4 \ti \widetilde{T}^4$ onto $T^4$ and $\widetilde{T}^4$. 
We summarize on the Poincar\'e line bundle as follows.
\begin{eqnarray*}
\ba{ccccc}
~   &       ~            &     \cP        &            ~             &  ~      \\
~   &       ~            &     \dar       &            ~             &  ~      \\
~   &       ~            & T^4 \ti \widetilde{T}^4 &            ~             &  ~      \\
~   & \st{\pi}{\swarrow} &       ~        & \st{\tilde{\pi}}{\searrow} &  ~      \\
T^4 &       ~            &       ~        &            ~             &  \widetilde{T}^4
\ea
\end{eqnarray*}

\subsection{Nahm Transformation}

Let $E \rightarrow T^4$ be a $U(N)$ vector bundle with the 2nd Chern number $c_2(E) = k$ 
and $A$ be a $U(N)$ gauge potential on $E$.   
The tensor product bundle $\pi^* E \otimes \cP$ is the $U(N)$ vector bundle over $T^4 \times \widetilde{T}^4$ 
equipped with the gauge potential $A \otimes 1_{[1]} + 1_{[N]} \otimes \omega$.
Taking the slice along $T^4 \times \{ \x \}$, 
we regard $\pi^*E \otimes \cP \vert_{T^4 \ti \left\{ \x \right\} } \rightarrow T^4 \times \{ \x \}$ 
as a $U(N)$ vector bundle over $T^4$. In particular, 
the gauge potential on $\pi^*E \otimes \cP \vert_{T^4 \ti \left\{ \x \right\} }$ is given by 
$A_{\x} = A \otimes 1_{[1]} + 1_{[N]} \otimes i \x_{\mu} dx^{\mu}$. 
The field strength $F_\x$ of $A_\x$ equals to $F$ of $A$.

%%where $E$ is a complex vector bundle on $T^4$ with a Hermitian metric and $G=U(N),~C_2=k$. 
%%First we pull the bundle $E$ back by the projection $\pi$. 
%%The gauge field on $\cP\ot \pi^*E\vert_{T^4\ti\left\{\x\right\}}$ is defined by 
%%$A_\x:=A\ot 1_{\scr \cL}+1_{[N]}\ot i\x_\mu dx^\mu$. 
%%The field strength $F_\x$ of $A_\x$ equals to $F$ of $A$. 
%%The covariant derivative of $A_\x$ is denoted by $D[A_\x]:=d+A_\x$. 

We introduce the Dirac operator. Let $S^{\pm} \rar T^4$ be a spinor bundle on $T^4$.
The Dirac operator acting on the sections 
$\Gamma( T^4, S^\pm \ot \pi^*E \otimes \cP \vert_{T^4 \ti \left\{ \x \right\} } )$ is given by
\begin{eqnarray}
\cD  [ A_\x ] &=&   e_\mu \ot ( \del_\mu + A_\mu + i\x_\mu),\nn
\cDb [ A_\x ] &=& \eb_\mu \ot ( \del_\mu + A_\mu + i\x_\mu ),
\end{eqnarray}
where $e_\mu = ( -i\sigma_a, 1_2),~\eb_\mu = ( i\sigma_a, 1_2)$
and  $\sigma_a~(a=1,2,3)$ are the Pauli matrices: % $\sigma_i$ are defined as
\begin{eqnarray}
%\mbox{The Pauli matrices}~~~
\sigma_1=
\left(\begin{array}{cc}
0 & 1\\
1 & 0
	 \end{array}\right),~
\sigma_2=\left(\begin{array}{cc}
0 & -i\\
i & 0
	 \end{array}\right),~
\sigma_3=\left(\begin{array}{cc}
1 & 0\\
0 & -1
	 \end{array}\right).
\end{eqnarray} 
These are the {\it Weyl operators} rather than the Dirac operators, 
however, we use the word the ``Dirac operator'' for simplicity.

\begin{figure}[htb]
\hspace{2cm}
\begin{center}
\includegraphics[width=80mm]{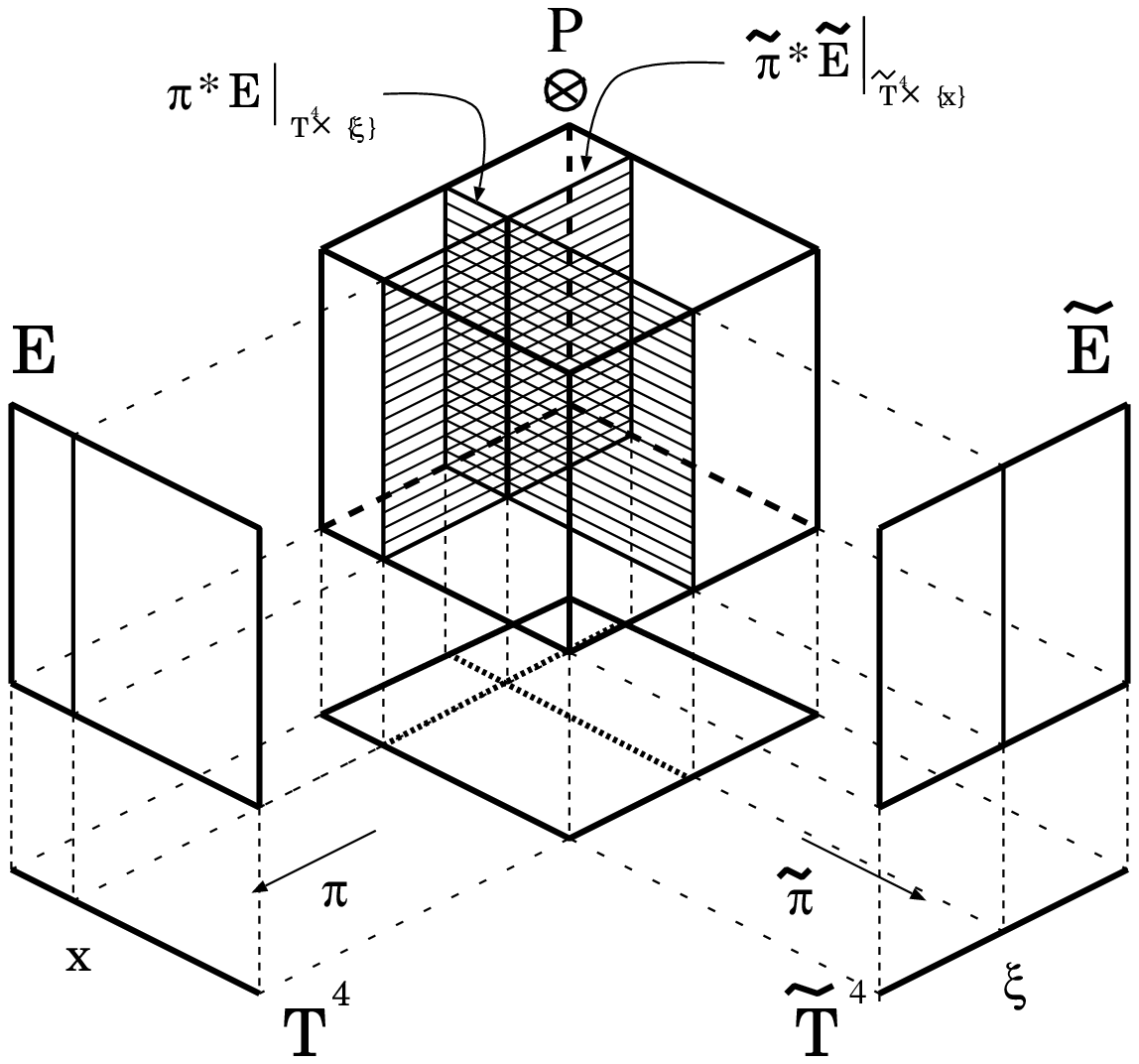}
\caption{The stage of the Nahm transformation}
\label{Poin}
\end{center}
\end{figure}

The dual vector bundle $\widetilde{E} \rightarrow \widetilde{T}^4$ 
can be constructed by using the Dirac zero modes.  
%%Here let us construct the dual vector bundle $\widetilde{E}$ on $\widetilde{T}^4$ 
%%by using the Dirac zero-mode $\psi_\x^p(x),~p=1,\cdots,k$.
The fiber $\widetilde{E}_\x$ is identified with $\Ker \cDb[A_\x]$. 
Since the Atiyah-Singer family index theorem implies $\dim \Ker \cDb [A_\x]=k$, the rank of $\widetilde{E}$ is $k$. 
Actually, $\widetilde{E}$ is a $U(k)$ vector bundle equiped with the dual $U(k)$ gauge potential $\widetilde{A}$. 
For the description of the dual gauge potential, 
we regard $\widetilde{E}$ as a sub-bundle of $\widetilde{H}$, 
where $\widetilde{H} \rar \widetilde{T}^4$ is an infinite-dimensional trivial vector bundle 
with the fiber $\widetilde{H}_\x = L^2(T^4, S^+ \ot \pi^* E \ot \cP \vert_{T^4 \ti \left\{ \x \right\} } )$. 
%%$\widetilde{E}$ is a sub-bundle of $\widetilde{H}$, 
%%since the fibre $\widetilde{E}_\x = \Ker \cDb [A_\x]$ is the finite dimensional subspace of $\widetilde{H}_\x$.  
(See Fig. \ref{Poin}.) 
By using the projection $P : \widetilde{H} \rar \widetilde{E}$, 
diffrential $\widetilde{d}$ on $\Gamma ( \widetilde{T}^4, \widetilde{H} )$ induces 
the covariant derivative $\widetilde{D} = \widetilde{d} + \widetilde{A}$ as 
\begin{eqnarray}
\lab{covder}
\widetilde{D} = P \widetilde{d} : \Gamma( \widetilde{T}^4, \widetilde{E}) \rar \Gamma( \widetilde{T}^4, \Lambda^1 \ot \widetilde{E} ).
\end{eqnarray}

The $U(k)$ gauge potential $\widetilde{A}$ is expressed by using the Dirac zero modes as 
\begin{eqnarray}
\label{gaugeba}
\widetilde{A}_\mu^{pq} ( \x ) 
= \int_{T^4} d^4x ~\psi^{p \dagger}_{\x}( x ) \fr{\del}{\del \x^\mu} \psi^q_{\x}( x )
\end{eqnarray}
where $\psi_{\x}^p~(p=1,2,\ldots,k)$ are the normalized Dirac zero-modes. 
\begin{eqnarray}
\cDb [ A_\x ] \psi_{\x}^p( x ) = 0, 
~~\int_{T^4 }d^4x ~\psi^{p \dagger}_{\x}( x ) \psi^q_{\x}( x ) = \delta^{p q}. 
\end{eqnarray}

%\begin{figure}[htbn]
%\begin{center}
%\epsfile{file=poincare1.eps,height=10cm}
%\epsfxsize=100mm
%\hspace{0cm}
%\epsffile{poincare1.eps}
%\caption{The stage of Nahm transformation}
%\label{Poin}
%\end{center}
%\end{figure}

%%Now let us introduce the projection 
%%\begin{eqnarray}
%%P : \widetilde{H} \rar \widetilde{E}
%%\end{eqnarray}
%%and define the covariant derivative as follows
%%\begin{eqnarray}
%%\lab{covder}
%%\widetilde{D}=P\widetilde{d}:
%%\Gamma(\widetilde{T}^4,\widetilde{E})\rar\Gamma(\widetilde{T}^4,\Lambda^1\ot\widetilde{E}).
%%\end{eqnarray}

%%Then gauge fields $\widetilde{A}$ on $\widetilde{E}$ are specified. 
%%This is the Nahm transformation: $\cN:(E,A)\map (\widetilde{E},\widetilde{A})$.
%%The concrete representation of the dual gauge fields is:
%%\begin{eqnarray}
%%\label{gaugeba}
%%\widetilde{A}_\mu^{pq}=\int_{T^4}d^4 x~\psi^{\dagger p}\fr{\del}{\del \x^\mu}\psi^q,
%%\end{eqnarray}
%%where $\psi^p~(p=1,2,\ldots,k)$ is the $k$ normalizable Dirac zero-modes. 

The 2nd Chern number of $\widetilde{E}$ turns out to be $N$ \cite{BrvB, Schenk}. 
Thus we obtain the $U(k)$ vector bundle $\widetilde{E} \rightarrow \widetilde{T}^4$ 
with $c_2(\widetilde{E}) = N$ and the $U(k)$ gauge potential $\widetilde{A}$ on $\widetilde{E}$. 
This gives  
the Nahm transformation $\cN:(E,A)\map (\widetilde{E},\widetilde{A})$.
\begin{eqnarray*}
\ba{ccccccc}
~&~&~&~&\pi^*E\ot\cP&~&~\\
~&~&~&~&\dar&~&~\\
~&~&E&~&T^4\ti\widetilde{T}^4&~&\widetilde{E}\\
~&~&\dar&\st{\pi}{\swarrow}&~&\st{\tilde{\pi}}{\searrow}&\dar\\
~&~&T^4&~&~&~&\widetilde{T}^4
\ea
\end{eqnarray*}

%%The similar argument is possible from $\widetilde{T}^4$ 
The inverse transformation is carried out by following the same procedure. 
Let $\widetilde{E} \rightarrow \widetilde{T}^4$ be a $U(k)$ vector bundle with $c_2(\widetilde{E}) = N$ 
and $\widetilde{A}$ be a $U(k)$ gauge potential on $\widetilde{E}$.   
We regard the slice of the tensor product bundle $\widetilde{\pi}^* \widetilde{E} \otimes \cP$ 
along $\{ x \} \times \widetilde{T}^4$ as a $U(k)$ vector bundle over $\widetilde{T}^4$ 
equipped with the gauge potential 
$\widetilde{A}_{x} = \widetilde{A} \otimes 1_{[1]} - 1_{[k]} \otimes i x^{\mu} d\x_{\mu}$. 
The field strength $\widetilde{F}_x$ of $\widetilde{A}_x$ equals to $\widetilde{F}$ of $\widetilde{A}$.
The Dirac operator acting on the sections 
$\Gamma( \widetilde{T}^4, \widetilde{S}^\pm \ot \pi^* \widetilde{E} \otimes \cP \vert_{ \{ x \} \ti \widetilde{T}^4} )$ 
takes the form 
\begin{eqnarray}
\widetilde{\cD} {[ \widetilde{A}_x] } &=& e_\mu \ot ( \widetilde{\del}^\mu + \widetilde{A}^\mu - ix^\mu ),\nn
\widetilde{\cDb}{[ \widetilde{A}_x] } &=& \eb_\mu \ot ( \widetilde{\del}^\mu + \widetilde{A}^\mu - ix^\mu).
\end{eqnarray}
By using the zero modes of $\widetilde{\cDb}{[ \widetilde{A}_x] }$, consisting of $N$ normalizable ones,   
we obtain a $U(N)$ vector bundle $E \rightarrow T^4$ with $c_2( E ) = k$ and a $U(N)$ gauge potential $A$ on $E$. 
This gives the inverse Nahm transformation $\widetilde{\cN}:(\widetilde{E},\widetilde{A})\map(E,A)$.
\begin{eqnarray*}
\ba{ccccccc}
~ & ~ & ~    & ~                  & \tilde{\pi}^*\widetilde{E} \ot \cP & ~                        & ~ \\
~ & ~ & ~    & ~                  & \dar                           & ~                        & ~ \\
~ & ~ & E    & ~                  & T^4 \ti \widetilde{T}^4                 & ~                        & \widetilde{E} \\
~ & ~ & \dar & \st{\pi}{\swarrow} & ~                              & \st{\tilde{\pi}}{\searrow} & \dar\\
~ & ~ & T^4  & ~                  & ~                              & ~                        &  \widetilde{T}^4
\ea
\end{eqnarray*}

It can be shown  $\cN\widetilde{\cN}=\widetilde{\cN}\cN=id.$ \cite{BrvB, Schenk}. 
Thus, eventually, $\widetilde{\cN}$ turns out to be the inverse of $\cN$. 
The Nahm duality is summarized in the following dictionary. 
%%We can prove that the Nahm transformation is a one-to-one map,
%%that is, $\cN\widetilde{\cN}=$id. and  $\widetilde{\cN}\cN=$id.
%%Summary is the following:
%\vs
%\unl{\bf Nahm transformation}
\begin{eqnarray*}
\lab{summary}
\ba{ccc}
%E&~&\widetilde{E}\\
%\dar&~&\dar\\
%T^4&~&\widetilde{T}^4\\
G=U(N)&~&\widetilde{G}=U(k)\\
k\mbox{-instanton on } T^4&\st{1~:~1}{\longlr}
&N\mbox{-instanton on } \widetilde{T}^4\\
~(\mbox{coord. } x^\mu) 
&~&~(\mbox{coord. }\xi^\mu)\\
%~&~&~\\
~&\mbox{massless Dirac eq.}&~\\
~&\cDb \psi =0&~\\
\mbox{instanton}~:~A_{\mu[N]}&\st{k \mbox{ solutions} :~\psi({\x},x)}{\longr}&
\dis\widetilde{A}_{\mu[k]}=\int_{T^4}d^4 x~\psi^\dagger\fr{\del}{\del \x^\mu}
\psi\\
~&~&~\\
~&\mbox{massless Dirac eq.}&~\\
~&\widetilde{\cDb}v=0&~\\
\dis A_{\mu[N]}=\int_{\widetilde{T}^4}d^4 \x~v^\dagger\fr{\del}{\del x^\mu}
v
&\st{N \mbox{ solutions}: 
~v(x,\x)}{\longl}&\mbox{instanton}~:~\widetilde{A}_{\mu[k]}\\
\ea
\end{eqnarray*}

%D-brane interpretation

\subsection{A Derivation of the ADHM construction from the Nahm transformation}

We can understand the duality in the ADHM/Nahm construction in some limit of the Nahm duality.  

\begin{itemize}
\item Taking all four radii of the four-torus infinity $\Rightarrow$ ADHM construction of instantons

Then the radii of the dual torus become zero.
Hence the dual torus shrinks into one point and derivatives become meaningless because
derivatives measure difference between two points.
As the result, all derivatives in the dual ASD equation and the dual massless Dirac equation drop out and 
the differential equations becomes matrix equations. This naively yields the ADHM duality:
one-to-one correspondence between the moduli space of the ASD equations on $\mathbb{R}^4$ (=infinite-size torus) 
and the moduli space of the matrix equations. For more detailed discussion, see \cite{vanBaal}.

\item Taking three radii infinity and the other radius zero $\Rightarrow$ Nahm construction of monopoles

\end{itemize}

\section{ADHM Construction of Instantons}

In this section, 
we explain the ADHM construction of instantons on noncommutative Euclidean $\mathbb{R}^4$. 
Let $B_1$ and $B_2$ be $k \times k$ complex matrices, 
and $I$ and $J$ be $k \times N$ and $N \times k$ complex matrices. 
The ADHM data for noncommutative $U(N)$ $k$-instantons are quadruple matrices 
$B_1, B_2, I$ and $J$ which satisfy the noncommutative 
version of the ADHM equation.  
\begin{eqnarray}
\label{adhm_now}
\mu_{ \mathbb{R} } 
\equiv {[ B_1, B_1^\dagger ]} + [ B_2, B_2^\dagger ]  + I I^\dagger - J^\dagger J 
= \zeta,
&&
\mu_{ \mathbb{C} } 
\equiv {[ B_1, B_2 ]} + I J = 0,
\end{eqnarray}
where 
$\zeta \equiv  
- [ \hat{z}_1, \hat{\bar{z}}_1 ] - [ \hat{z}_2, \hat{\bar{z}}_2 ] 
= -2( \theta_1 + \theta_2 )$  
in the first equation of (\ref{adhm_now}) is a non-negative constant 
originated in the noncommutativity (\ref{nc_coord}). 
In the case that $\theta^{\mu \nu}$ is anti-self-dual, 
$\zeta$ vanishes and the noncommutative ADHM equation (\ref{adhm_now}) 
coincides with the commutative ADHM equation. 
With the ADHM data, we associate ``0-dimensional Dirac equation'' 
\begin{eqnarray}
\label{nc_0dirac}
\hat{\nabla}^\dagger \cdot \hat{V} = 0,~~~
\hat{\nabla} (x) = \left( \begin{array}{cc} 
                             I^\dagger       & J \\
               \hat{\bar{z}}_2 - B_2^\dagger & ~-( \hat{z}_1 - B_1 ) \\
               \hat{\bar{z}}_1 - B_1^\dagger & ~\hat{z}_2 - B_2
                    \end{array} \right). 
\end{eqnarray}
The solution $\hat{V}$ of equation (\ref{nc_0dirac}) is an 
$(N+2k) \times N$ matrix.  

In the noncommutative case, 
we need to take care about the existence of zero-modes of $\hat{V}$. 
We may normalize $\hat{V}$ so that 
\begin{eqnarray}
\label{0norm}
\hat{V}^\dagger \cdot \hat{V} = 1_N.  
\end{eqnarray}
By using the normalized $\hat{V}$, we finally obtain the corresponding instanton solution as
\begin{eqnarray}
\label{nc_4inst} 
\hat{A}_\mu = \hat{V}^\dagger \cdot \partial_\mu \hat{V}. 
\end{eqnarray}
Actually, owing to the construction similar to the commutative case, 
the gauge potential (\ref{nc_4inst}) turns out to satisfy 
the noncommutative anti-self-dual Yang-Mills equation (\ref{bps_instanton}).
The instanton number is computed to be $k$.

We will focus mainly on the case of $(N, k) = (2, 1)$ in the subsequent discussions.  
For further reviews on noncommutative instantons, 
see \cite{Furuuchi3, Hamanaka3, KoSc2, Lechtenfeld, Nekrasov2, Schaposnik}.

\subsection{ADHM Construction of Commutative Instantons}

To argue the ADHM construction of noncommutative instantons, 
it is convenient to recall the commutative case. 
We illustrate the case by the Belavin-Polyakov-Schwartz-Tyupkin (BPST) instanton \cite{BPST}, 
that is, a $SU(2)$ 1-instanton on commutative Euclidean $\mathbb{R}^4$.

In the case of $k = 1$, 
$B_1$ and $B_2$ can be chosen as arbitrary complex numbers. 
The ADHM equation is then solved by using a real non-negative number $\rho$ as 
\begin{eqnarray}
\label{BIJ_BPST}
B_1 = \alpha_1, ~B_2 = \alpha_2, ~I = ( \rho, \,0 ),
~J = \left(\begin{array}{c}
                   0   \\[-1mm]
                  \rho
     \end{array} \right),
~~~\alpha_{1, 2} \in \C, ~\rho \in \R_{\geq 0}.
\end{eqnarray} 
The complex numbers $\alpha_{i = 1, 2}$ describe the position of the instanton.  
They can be always made zero by the coordinate shift 
$z_{i} \rightarrow z_{i} + \alpha_{i}$ in the Dirac equation. 
We will put $\alpha_1 = \alpha_2 = 0$, without any loss.
The normalized solution of the Dirac equation (\ref{nc_0dirac}) takes the form
\begin{eqnarray}
\label{V_BPST}
V = \frac{1}{\sqrt{\phi}} 
\left( \begin{array}{cc} 
        z_2    &    z_1     \\
    -\bar{z}_1 & \bar{z}_2  \\
      -\rho    &      0     \\
         0     &   -\rho    \\
\end{array}\right),
~~~\phi = r^2 + \rho^2, ~~r = \sqrt{ {z}_1 \bar{z}_1 + {z}_2 \bar{z}_2 }. 
\end{eqnarray}
The corresponding instanton solution 
$\displaystyle A_\mu = V^\dagger \del_\mu V$ reads
\begin{eqnarray}
\lab{bpstkai}
A_\mu = \frac{r^2}{r^2 + \rho^2} g^{-1} \partial_\mu g,~~~
g = \frac{1}{r}
      \left( \begin{array}{cc} 
                z_2 & z_1 \\
         -\bar{z}_1 & \bar{z}_2
             \end{array} \right).
\end{eqnarray}
The field strength is computed to be 
\begin{eqnarray}
\label{F_BPST}
F_\mn 
= \fr{ 2i \rho^2 }{( r^2 + \rho^2 )^2 } \eta_{\mn},
\end{eqnarray} 
where $\eta_{\mu \nu}$ is a $2 \times 2$ matrix 
obtained by contracting 't Hooft's anti-self-dual eta symbol 
with the Pauli matrices as $\eta_\mn = \eta^{a}_\mn \sigma_a$. 
This is exactly the BPST instanton solution \cite{BPST}. 
$\rho$ describes the size of the instanton.

\begin{figure}[htbn]
\begin{center}
\hspace{1cm}
\includegraphics[width=9cm]{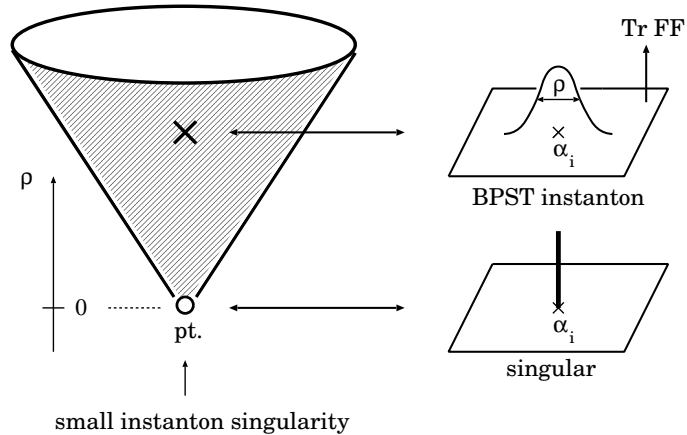}
\caption{Instanton moduli space %${\cal M}_0$
and configurations of the BPST instanton ($\zeta=0$)}
\label{bpst_com}
\end{center}
\end{figure}

In the limit $\rho \rightarrow 0$, 
the field strength (\ref{F_BPST}) concentrates at the origin and becomes singular. 
In the framed instanton moduli space ${\cal{M}}$,  
it is a singularity called the small instanton singularity. 
The moduli space is depicted in Fig. \ref{bpst_com}, 
where the distributions of $\mbox{Tr}F_{\mn}F^{\mn}$ 
for the cases of $\rho \neq 0$ and $\rho = 0$ 
are illustrated on the right-hand side.

\vspace{3mm}

\subsection{ADHM Construction of NC instantons ($\zeta>0$)}

Moduli space of noncommutative instantons depends on the constant $\zeta$ \cite{Nakajima2, Nakajima3}. 
When $\zeta = 0$, the moduli space coincides with that of commutative instantons 
and contains the small instanton singularities. On the other hand, when $\zeta > 0$, 
such singularities disappear and are replaced by a new class of smooth instantons, $U(1)$ instantons.  
In this subsection, we examine the case of $\zeta >0$. Let $\theta_{1} = \theta_{2} = - \zeta/4$.   
We construct a noncommutative instanton located at the origin,  
and observe how the new class of instantons appears on the noncommutative space,  
replacing the small instanton singularity.

\subsubsection{NC $U(1)$ instantons ($\zeta>0$)}

It is convenient to consider first the case of noncommutative $U(1)$ instantons. 
The ADHM data for a noncommutative $U(1)$ $1$-instanton is 
\begin{eqnarray}
\label{ADHM_(1,1)} 
B_1 = B_2 = 0, 
~I = \sqrt{\zeta}, ~J = 0.
\end{eqnarray} 
The Dirac equation (\ref{nc_0dirac}) reads  
\begin{eqnarray}
\label{eq_V_(1,1)}
\hat{\nabla}^\dagger \cdot \hat{V} 
= \left( \begin{array}{ccc}
         \sqrt{\zeta} & \hat{z}_2        & \hat{z}_1 \\
         0            & -\hat{\bar{z}}_1 & \hat{\bar{z}}_2 
       \end{array} \right) 
\cdot \hat{V}
= 0. 
\end{eqnarray}
The following form of $\hat{V}$ gives a solution of (\ref{eq_V_(1,1)}).  
\begin{eqnarray}
\label{sol_0dirac_0}
\hat{V}_1 
= \left( \begin{array}{cc}
             \hat{z}_1 \hat{\bar{z}}_1 + \hat{z}_2 \hat{\bar{z}}_2  \\
                       - \sqrt{\zeta} \hat{\bar{z}}_2               \\
                       - \sqrt{\zeta} \hat{\bar{z}}_1
       \end{array} \right), 
\end{eqnarray}
where $\hat{V}_1$ is an operator acting on the Fock space $\cal{H}$. 
Since $2 ( \theta_1 + \theta_2 ) = - \zeta < 0$, 
the commutation relations (\ref{nc_cpx_coord}) imply that 
at least either coordinates $\hat{\bar{z}}_1$ or $\hat{\bar{z}}_2$ 
is the annihilation operator. 
Actually, at present, we put $ \theta_1 = \theta_2 = - \zeta/4$. 
This means that both the coordinates are the annihilation operators.  
$\hat{V}_1$ has the kernel, an operator zero-mode $\vert 0, 0 \rangle$. 
Thus, it can not provide a normalized solution to satisfy the condition (\ref{0norm}).

K. Furuuchi \cite{Furuuchi} shows that 
$\Vh_1$ yields a smooth anti-self-dual instanton solution on the subspace 
 $\cH_1 \equiv \cH \setminus \mathbb{C} \vert 0, 0 \ket$.  
Using the shift operator, he also discussed that 
it can be modified to satisfy the condition \eqref{0norm} on $\cH$. 
Let $\hat{U}_1$ be a shift operator of the form 
\begin{eqnarray}
\Uh_1 
= \sum_{n_1 = 1}^{+\infty} \sum_{n_2 = 0}^{+ \infty} \vert n_1, n_2 \ket \bra n_1, n_2 \vert 
+ \sum_{ n_2 = 0 }^{+\infty} \vert 0, n_2 \ket \bra 0, n_2 + 1 \vert. 
\label{4shift}
\end{eqnarray}
It satisfies $\Uh_1 \Uh_1^\dagger = 1$ and $\Uh_1^\dagger \Uh_1 = 1 - \hat{P}_1$, 
where $\hat{P}_1 = \vert 0, 0 \rangle \langle 0, 0 \vert$ is the projection operator to the kernel.
The modification reads \cite{Furuuchi2} 
\begin{eqnarray}
\Vh 
= \Vh_1 \hat{{\beta}}_1 \Uh_1^\dagger, 
&&
\hat{{\beta}}_1 
\equiv ( 1 - \Ph_1 ) \left( \Vh_1^{\dagger} \Vh_1 \right)^{ - \half } ( 1 - \Ph_1 ). 
\label{Vh_modified}
\end{eqnarray}
The modified operator (\ref{Vh_modified}) satisfies $\Vh^\dagger \Vh = 1$ on $\cH$. 
This can be seen by a direct computation of $\Vh^\dagger \Vh$  
using (\ref{sol_0dirac_0}) and (\ref{4shift}) together with $\hat{\beta}_1$ in the following form. 
\begin{eqnarray}
\hat{{\beta}}_1 
= \frac{2}{\zeta} 
\sum_{(n_1,n_2)\neq (0,0)}
\frac{ \vert n_1, n_2 \rangle \langle n_1, n_2 \vert}{\sqrt{ ( n_1 + n_2 )( n_1 + n_2 + 2 )}}. 
\end{eqnarray}

Thus the obtained normalized solution \eqref{Vh_modified}  gives in the sequel 
a smooth anti-self-dual instanton on $\cH$ with the 
instanton number equal to $1$.

\subsubsection{NC $U(2)$ instantons ($\zeta>0$)}

Noncommutative BPST solution is a noncommutative version of the BPST instanton 
and can be also obtained by  the ADHM construction.   
The ADHM data is 
\begin{eqnarray}
B_1 = B_2 = 0,
~~~I = ( \sqrt{\rho^2 + \zeta}, 0 ),
~~~J = \left( \begin{array}{c} 
                 0 \\ 
                 \rho 
              \end{array} \right). 
\label{BIJ_BPST_NC1}              
\end{eqnarray} 
Compared with the commutative case (\ref{BIJ_BPST}), 
$I$ in the ADHM data is deformed by $\zeta$. 
This implies that the size of instanton is not less than $\sqrt{\zeta}$.

Normalized solution of the Dirac equation can be found 
by Furuuchi's approach \cite{Furuuchi2, Furuuchi3}. 
We first see that the following form of $\hat{V}$ is a solution of the Dirac equation. 
\begin{eqnarray}
\hat{V}_1 =  \Bigl( \hat{V}_1^{(1)}, \,\hat{V}_1^{(2)} \Bigr) 
\label{hat_V_NC_U(2)_1}
\end{eqnarray}
where 
\begin{eqnarray}
\hat{V}_1^{(1)} 
= \left( \begin{array}{cc}
          \hat{z}_1 \hat{\bar{z}}_1 + \hat{z}_2 \hat{\bar{z}}_2 \\
                                0                               \\
             - \sqrt{ \rho^2 + \zeta } \,\hat{\bar{z}}_2        \\
             - \sqrt{ \rho^2 + \zeta } \,\hat{\bar{z}}_1
        \end{array}  \right), 
&&       
\hat{V}_1^{(2)} 
= \left( \begin{array}{cc}
                                0                               \\
  \hat{z}_1 \hat{\bar{z}}_1 + \hat{z}_2 \hat{\bar{z}}_2 + \zeta \\
                          \rho \,\hat{z}_1                      \\
                        - \rho \,\hat{z}_2
         \end{array}\right).
\label{component_hat_V_NC_U(2)_1}
\end{eqnarray}         
Apparently, $\hat{V}_1^{(1)}$ has the same kernel as (\ref{sol_0dirac_0}), 
while $\hat{V}_1^{(2)}$ is injective.
To obtain a normalized solution from (\ref{hat_V_NC_U(2)_1}), 
we introduce the shift operator $\hat{U}_1$ which satisfies 
\begin{eqnarray}
\Uh_1 \Uh_1^\dagger = 1,
&&
\Uh_1^\dagger \Uh_1 
= 1 - \left( \begin{array}{cc} \hat{P}_1 & 0 \\
                                   0 & 0 
             \end{array} \right). 
\label{condition_hat_U_NC_U(2)_1}             
\end{eqnarray} 
It is actually given by  
\begin{eqnarray}
\Uh_1 
= \sqrt{ \frac{2}{\zeta} }
\left( \begin{array}{cc}
            \hat{\bar{z}}_2 & -\hat{z}_1  \\
            \hat{\bar{z}}_1 &  \hat{z}_2
         \end{array} \right)
\sum_{n_1 = 0}^{+ \infty} \sum_{n_2 = 0}^{+\infty} 
\frac{ \vert n_1, n_2 \rangle \langle n_1, n_2 \vert}{\sqrt{n_1 + n_2 + 1}}. 
\label{hat_U_NC_U(2)_1} 
\end{eqnarray} 
Combining these operators, 
the normalized solution $\hat{V} = ( \hat{V}^{(1)}, \hat{V}^{(2)} )$ 
is obtained in the following form. 
\begin{eqnarray}
\hat{V} 
= \hat{V}_1 
\cdot  \left( \begin{array}{cc} 
             \hat{\beta}_1^{(1)} &    0               \\[-1mm]
                        0        & \hat{\beta}_1^{(2)} 
              \end{array} \right) 
\cdot \hat{U}_1^\dagger,   
\label{hat_V_NC_U(2)}
\end{eqnarray}
where  
\begin{eqnarray}
&&
{\displaystyle \hat{\beta}_1^{(1)} 
= \frac{2}{\zeta} \sum_{( n_1, n_2 ) \neq ( 0, 0 )}
   \frac{ \vert n_1, n_2 \rangle \langle n_1, n_2 \vert }
            { \sqrt{ ( n_1 + n_2 )( n_1 + n_2 + 2 + \frac{2 \rho^2}{\zeta} )} }}, 
\nonumber \\
&&
{\displaystyle \hat{\beta}_1^{(2)} 
= \frac{2}{\zeta} \sum_{n_1 = 0}^{+ \infty} \sum_{ n_2 = 0}^{+\infty} 
      \frac{ \vert n_1, n_2 \rangle \langle n_1, n_2 \vert }
            { \sqrt{ ( n_1 + n_2 + 2 )( n_1 + n_2 + 2 + \frac{2 \rho^2}{\zeta} )} }}. 
\label{hat_beta_NV_U(2)_1}
\end{eqnarray}

\begin{figure}[htbn]
\begin{center}
\hspace{1cm}
\includegraphics[width=9cm]{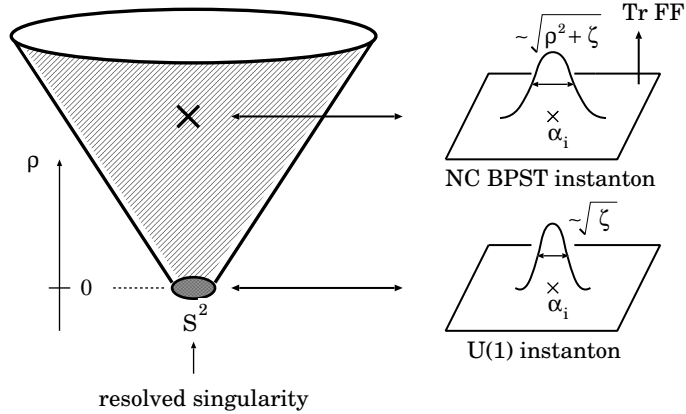}
\caption{Instanton moduli space %${\cal M}_\zeta$
and the configuration of the NC BPST instanton ($\zeta>0$)}
\label{NCbpst2}
\end{center}
\end{figure}

The normalized solution (\ref{hat_V_NC_U(2)}) turns out to 
give a smooth anti-self-dual instanton 
on $\cH$ with the instanton number equal to $1$, 
that is, the noncommutative BPST instanton. 
It keeps to be smooth in the limit $\rho \rar 0$ and becomes the $U(1)$ instanton. 
This is because $\Vh^{(2)}$ decouples at this limit and does not contribute to the field strength. (See Fig. \ref{NCbpst2}.)

%\vspace{3mm}

The BPST instantons on commutative and noncommutative spaces 
are summarized in the following table.

%%%%%%%%%%%%%%%%%%%
\begin{center}
\begin{tabular}{|c|c|c|} \hline
BPST instanton&$(N,k)=(2,1)$
&NC BPST instanton\\ \hline\hline
$\mu_{\mathbb{R}}=0,~\mu_{\mathbb{C}}=0$&
ADHM equation&$\mu_{\mathbb{R}}=\zeta>0 ,
~\mu_{\mathbb{C}}=0$
\\\hline
$B_{1,2}=\alpha_{1,2},$&ADHM data
&$B_{1,2}=\alpha_{1,2},$\\
$I=(\rho,0),J^t=(0,\rho)$&~&$I=(\sqrt{\rho^2+\zeta},0),J^t=(0,\rho)$
 \\\hline
$\mathbb{R}^4\times$ orbifold $\mathbb{C}^2/\mathbb{Z}_2$& moduli
space&$\mathbb{R}^4\times$ Eguchi-Hanson
	 $\widetilde{\mathbb{C}^2/\mathbb{Z}_2}$\\
(singular)&~&(regular) \\\hline
$F_{\mn}\rar$ (singular)
&zero-size limit
& $F_{\mn}\rar$ $U(1)$ instanton (regular)\\
\hline
\end{tabular}
\end{center}
%%%%%%%%%%%%%%%%%%%%

\subsection{ADHM Construction of NC instantons ($\zeta=0$)}

Finally we discuss the ADHM construction of noncommutative instantons 
at the special value, $\zeta=0$, 
where the noncommutative parameter is anti-self-dual 
and satisfies in the canonical form, $\theta_1 = - \theta_2 \neq 0$.  
We set $\theta_{1} = -\theta_{2} \equiv \theta > 0$. 
Hence the coordinate $\hat{\bar{z}}_1$ is now the creation operator, 
while $\hat{\bar{z}}_2$ is still the annihilation operator. 
%As a consequence, the situation 
%is changed when we solve the Dirac equation \eqref{nc_0dirac}.
%have no operator zero-mode. 
We consider the instantons at the origin.

\subsubsection{NC $U(1)$ instantons ($\zeta=0$)}

The instanton moduli space is the same as the commutative one 
and contains the small instanton singularities.
Let us first examine the $U(1)$ instanton that gives the small instanton singularity.
The ADHM data is trivial. 
\begin{eqnarray}
\label{ADHM_(1,1,ASD)} 
B_{1,2} = 0, ~I = 0, ~J = 0.
\end{eqnarray} 
The normalized solution of the Dirac equation (\ref{nc_0dirac}) is given \cite{Hamanaka} by 
\begin{eqnarray}
\label{v_instanton}
\Vh=
\left( \ba{c}
\Uh_1\\
0\\
\hat{P}_1
\ea\right),
\end{eqnarray}
where $\Uh_1$ is the shift operator in \eqref{4shift}. 
The normalization condition (\ref{0norm}) can be seen as 
$\displaystyle \hat{V}^\dagger \hat{V} = \hat{U}_1^\dagger \hat{U}_1 + \hat{P}_1 
= (1 - \hat{P}_1) + \hat{P}_1 = 1$.
Operators of covariant derivative and field strength take the following forms.
\begin{eqnarray}
\Dh_{z_i} = \Uh_1^\dagger \delh_{z_i} \Uh_1, 
~~~~
\hat{F}_{z_1 \bar{z}_1} = -\hat{F}_{z_2 \bar{z}_2} = \fr{1}{2\theta} \hat{P}_1.
\end{eqnarray}
The second Chern class  $(-1/16\pi^2) \mbox{Tr}\hat{F}_{\mn}\hat{F}^{\mn}$
is easily computed to be $(2\pi\theta)^{-2} \hat{P}_1$.
It is converted to the Gaussian distribution 
$(\pi\theta)^{-2}\exp(-r^2/\theta)$ in the star-product formalism 
by the inverse Weyl-transformation.
The instanton number is found to be $1$.

This type of exact solutions is first given in \cite{AGMS}
and plays important roles in string theory (e.g. \cite{Harvey2}).

\subsubsection{NC $U(2)$ instantons ($\zeta=0$)}

When $\zeta=0$, 
the ADHM construction of noncommutative BPST instantons 
is simpler than the previous case. 
The ADHM data is the same as the commutative one \eqref{BIJ_BPST}.  
\begin{eqnarray}
B_{1,2} = 0,
~~~I = (\rho, 0 ),
~~~J = \left( \begin{array}{c} 
                 0 \\ 
                 \rho 
              \end{array} \right).
\end{eqnarray}
The normalized solution of the Dirac equation 
\eqref{nc_0dirac} is \cite{Furuuchi4}: 
\begin{eqnarray}
\label{hat_V_NC_U(2)_2}
\Vh\hspace{-2mm}
&=&
\hspace{-2mm}
(\Vh^{(1)},\Vh^{(2)}),\\
\Vh^{(1)}\hspace{-2mm}&=&
\hspace{-2mm}
\left(
\begin{array}{c}
\hat{z}_2\\
-\hat{\bar{z}}_1\\
-\rho\\
0
\end{array}
\right)\sum_{n_1,n_2}
\frac{\vert n_1,n_2\ket\bra n_1,n_2\vert}
{\sqrt{2\theta(n_1+n_2+1)+\rho^2}},~
\Vh^{(2)}=
\left(
\begin{array}{c}
\hat{z}_1\\
\hat{\bar{z}}_2\\
0\\
-\rho
\end{array}
\right)
\sum_{n_1,n_2}
\frac{\vert n_1,n_2\ket\bra n_1,n_2\vert}{\sqrt{2\theta(n_1+n_2)+\rho^2}}.\no
\end{eqnarray}
This possesses similar structure as the commutative one \eqref{V_BPST} 
($r^2\leftrightarrow 2\theta(n_1+n_2) $).
The solution \eqref{hat_V_NC_U(2)_2} gives a smooth anti-self-dual instanton 
with the instanton number $1$. In the limit $\rho \rar 0$, 
this reduces to that of the $U(1)$ 1-instanton \eqref{v_instanton}.

\section{Conclusion and Discussion}

In this paper, we constructed exact noncommutative
instanton solutions by the ADHM procedure.
We further prove one-to-one correspondence between
the moduli space of the $U(N)$ $k$-instantons 
and the moduli space of the ADHM data labeled by $(N,k)$ \cite{HaNa}.
In the proof, we apply Furuuchi's observation 
and several properties on noncommutative field theory 
(especially NC-deformed index theorem by Maeda and Sako \cite{MaSa})
to all the other ingredients of the ADHM construction and 
prove the existence of them in the operator sense.
%(In the star product formalism, perturbative discussion 
%with respect to the noncommutative parameter is made in \cite{MaSa}.)
As a result, we obtain the following formula
(a noncommutative version of the {\it Corrigan-Goddard-Osborn-Templeton
or Osborn formula} \cite{CGOT, Osborn2}) 
\begin{eqnarray}
\label{CGOT}
\int d^4 x {\mbox{ Tr}}_N ( F_{\mu\nu}\star F^{\mu\nu} ) 
= -\int d^4 x \partial^2 \partial_\mu ( {\mbox{Tr}}_k f^{-1} \star \partial^{\mu}f ),
~f := ( \nabla^\dagger \star \nabla )^{-1}.
\end{eqnarray}
By using this formula and the asymptotic behavior of 
$f\simeq \vert x \vert^{-2}1_k$, the instanton number (the second 
Chern number) is computed after surface integration as
\begin{eqnarray*}
\frac{-1}{16\pi^2}\int d^4 x {\mbox{ Tr}}_N ( F_{\mu\nu}\star F^{\mu\nu} ) 
= {\mbox{Tr}}_k 1_k=k.
\end{eqnarray*}
%which is crucial to show that any ADHM data labeled by $k$ 
%yields $k$-instanton solutions even for the $U(1)$ case. 
This directly shows an origin of the instanton number 
from the language of the ADHM data.
(For other discussion on the instanton number,  
%which has been discussed by several authors for explicit instanton solutions
see e.g. \cite{CKT, Furuuchi3, IKS, Sako, Tian, STZ}.)
Detailed proofs and other results are seen in our paper \cite{HaNa}.

%\section{Conclusion and Discussion}

\bigskip
\noindent
{\bf Acknowledgments}

The authors thank the Yukawa Institute for Theoretical Physics at Kyoto University. 
Discussions during the YITP workshop on ``Field Theory and String Theory'' (YITP-W-12-05) were useful to complete this work. 
The authors are also grateful to A.~Sako for fruitful discussion.
MH is supported in part by the Daiko Foundation,
the Toyoaki Scholarship Foundation, and 
the Grant-in-Aid for Young Scientists (\#23740182). 
TN is supported in part by the Grant-in-Aid for Scientific Research No. 24540223.  

%\section*{References}

\end{document}